# Metal insulator transition and magnetotransport anomalies in perovskite SrIr$_{0.5}$Ru$_{0.5}$O$_3$ thin films

Abhijit Biswas, Yong Woo Lee, Sang Woo Kim and Yoon Hee Jeong[*]

Department of Physics, POSTECH, Pohang, 790-784, South Korea

## Abstract

We investigated the nature of transport and magnetic properties in SrIr$_{0.5}$Ru$_{0.5}$O$_3$, (SIRO) which has characteristics intermediate between a correlated non-Fermi liquid state and an itinerant Fermi liquid state, by growing perovskite thin films on various substrates (e.g. SrTiO$_3$ (001), (LaAlO$_3$)$_{0.3}$(Sr$_2$TaAlO$_6$)$_{0.7}$ (001) and LaAlO$_3$ (001)). We observed systematic variation of underlying substrate dependent metal-to-insulator transition temperatures ($T_{MIT}$ ~ 80 K on SrTiO$_3$, ~ 90 K on (LaAlO$_3$)$_{0.3}$(Sr$_2$TaAlO$_6$)$_{0.7}$ and ~ 100 K on LaAlO$_3$) in resistivity. At temperature 300 K $\geq T \geq T_{MIT}$, SIRO is metallic and its resistivity follows a $T^{3/2}$ power law; whereas insulating nature at $T < T_{MIT}$ is due to the localization effect. Magnetoresistance (MR) measurement of SIRO on SrTiO$_3$ (001) shows negative MR at $T <$ 25 K and positive MR at $T >$ 25 K, with negative MR $\propto \sqrt{B}$ and positive MR $\propto$ B$^2$; consistent with the localized-to-normal transport crossover dynamics. Furthermore, observed spin glass like behavior of SIRO on SrTiO$_3$ (001) at $T <$ 25 K in the localized regime, validates the hypothesis that (Anderson) localization favors glassy ordering. These remarkable features provide a promising approach for future applications and of fundamental interest in oxide thin films.

# I. INTRODUCTION

Transition metal oxides (TMOs) forging transition from one electronic phase to another, has been long debated problem in condensed matter physics as it display a plethora of many unique phenomena including metal-insulator transition (MIT) in strongly-correlated systems, high temperature superconductivity in layered cuprates, and colossal magnetoresistance (CMR) in perovskite manganites.[1-3] From the prolonged list of transition metal oxides, specially perovskites, Ruddlesden-Popper (RP) phases of strontium iridates and strontium ruthanates, $Sr_{n+1}R_nO_{3n+1}$ ($R$ = Ir and Ru, and $n$ = 1, 2, …, ∞) have been investigated with great interest as it shows coupling between lattice deformation, orbital degree of freedom, Coulomb interactions ($U$) and spin-orbit coupling ($SOC$) that give rise to several phenomena, including rich phase diagrams of MITs, unconventional magnetic ordering and the anomalous Hall effect.[4-7] Among various phases of $Sr_{n+1}Ir_nO_{3n+1}$, one of the compounds, Ir based strong spin orbit coupled ($5d^5$) perovskite $SrIrO_3$ ($n$ = ∞), is a strongly-correlated *paramagnetic non-Fermi liquid metal* without any long-range magnetic ordering. Moreover $SrIrO_3$ has characteristics that are very close to the metal/insulator phase boundary. It has high resistivity $\rho$ ~ 2 mΩ·cm at room temperature (RT) with *positive magnetoresistance* (MR ~ +1%) up to lowest temperature measured till date.[8-10] In contrast, $SrRuO_3$ is a Ru-based ($4d^4$) itinerant *ferromagnetic bad metal* with Curie temperature $T_C$ ~ 150 K and resistivity $\rho$ ~ 200 μΩ·cm at room temperature.[7] Its $\rho$ increases linearly with $T$ at $T > T_C$, but has Fermi liquid behavior at $T < T_C$. $SrRuO_3$ has been investigated extensively from bulk to its ultra-thin film limit and has been used as bottom electrode for device physics. $SrRuO_3$ has *Stoner* type (itinerant) ferromagnetism that is related to the hybridization of Ru ($t_{2g}$)–O (2p) orbitals and is thus affected by lattice distortion.[7] Moreover at low $T$ SrRuO$_3$ shows anisotropic *negative* MR (∼-5%) which is quite large with respect to those of conventional ferromagnetic metals.

Recently, efforts have been devoted to manipulating the magnetic $T_C$ and to tuning the transport properties in SrRuO$_3$ by replacing 4$d$ Ru with 3$d$,4$d$ transition metal elements.[11-15] This replacement causes localized orbitals to hinder the hybridization of Ru orbitals and results in formation of spin clusters that affect both magnetism and transport. Recently, efforts have also been made to synthesize polycrystalline SrIrO$_3$ doped with 3$d$ transition metals.[16,17]

Now, one of the most discernible questions from electronic point of view is what would happen if we replace the 4$d$ Ru with higher quantum numbered 5$d$ based transition metal Ir or vice versa; let's replace half of them and stay in the intermediate region for the time being. The consequences of replacing the 4$d$ Ru with 5$d$-based transition metal Ir or vice versa might allow coexistence of two ground states (SrRuO$_3$ and SrIrO$_3$) and push the system into different state. In this standard picture in terms of microscopic nature, 5$d$ transition metals have wider $d$ orbitals than do 4$d$ transition metals, so replacing 4$d$ element with 5$d$ should increase the band width ($W$). This change might also affect the overall magnetism, because SrRuO$_3$ has strong ferromagnetism of itinerant nature (band magnetism). Additionally, replacing 4$d$ ruthenium ($z = 44$) with 5$d$ iridium ($z = 77$) should increase SOC because SOC $\propto z^4$. The difference in transport and magnetism between SrIrO$_3$ and SrRuO$_3$ results from their different magnitudes of the effective interaction ($U/W$). Replacing the Ir$^{4+}$ ions by Ru$^{4+}$ ions, which have smaller ionic radius than does Ir$^{4+}$, would change the *B-O-B* bond angle; this distortion would then modify the overlap between neighboring $d$ orbitals mediated by oxygen 2p orbitals, thereby changing $W$ and enabling control of $U/W$. Also an interesting observation would be how the MR behaves due to the coupling of $SrIrO_3$, in which MR is positive, with $SrRuO_3$, in which MR is negative. If the MR of the compound material undergoes switching, it might be useful in memory device applications. Also for application purpose, extending the list of bottom electrodes having structural compatibility with perovskites and perovskite related structures is useful for the creation of new functional oxide electronic devices.

The purpose of this work is to show that the perovskite thin film $SrIr_{0.5}Ru_{0.5}O_3$; which has characteristics intermediate between a correlated non-Fermi liquid state and an itinerant Fermi liquid state, favors correlated state from resistivity point of view whereas magnetoresistance and magnetization measurement shows close resemblance to an itinerant electronic state.

## II. EXPERIMENTAL METHODS

We used pulsed laser deposition (KrF laser; $\lambda = 248$ nm) to grow $SrIr_{0.5}Ru_{0.5}O_3$ (SIRO) thin films from a polycrystalline target synthesized from a stoichiometric mixture of $SrCO_3$, $IrO_2$ and $RuO_2$ powder by a solid state reaction method. In general, during thin film growth metastable phases can be stabilized, and key parameters such as band width, SOC, and correlation of the system can be tuned by external parameters by using lattice-mismatched substrates.[18,19] We have attempted and succeeded to grow perovskite $SrIr_{0.5}Ru_{0.5}O_3$ thin films (~40 nm) on lattice-mismatched substrates ($SrTiO_3$ (001), $(LaAlO_3)_{0.3}(Sr_2TaAlO_6)_{0.7}$ (001) and $LaAlO_3$ (001)). The laser was operated at frequency 4 Hz, the substrate temperature was 680 °C, and oxygen partial pressure was 20 mTorr. After growth, all films were cooled down slowly (~7° C/min) in same oxygen partial pressure to compensate for any oxygen deficiency. All the substrates, SrTiO$_3$ (001), (LaAlO$_3$)$_{0.3}$(Sr$_2$TaAlO$_6$)$_{0.7}$ (001) and LaAlO$_3$ (001) crystals, were treated prior to growth to produce atomically flat surfaces.[20,21] (Henceforth these substrates will be designated as STO, LSAT and LAO). X-ray diffraction (XRD) measurements were performed by the Empyrean XRD System from PANalytical. XE-100 Advanced Scanning Probe Microscope was used for surface topography. To check elemental distributions, we used secondary ions mass spectrometry (SIMS) with a primary beam source of O$_2^+$ with impact energy of 7.5 keV. Electrical transport measurements were performed using the four-probe van der Pauw geometry within 10 K $\leq T \leq$ 300 K range.

### III. RESULTS AND DISCUSSION

At ambient pressure, SrIrO$_3$ and SrRuO$_3$ have different crystal structures. SrIrO$_3$ is of orthorhombic structure only at high pressure with a lattice constant of $a_c \sim$ 3.96 Å [22], whereas SrRuO$_3$, has a GdFeO$_3$ type distorted orthorhombic structure with a lattice constant of $a_c \sim$ 3.93 Å.[23] Assuming that SIRO should be pseudo-cubic with 3.93 Å $\leq a_c \leq$ 3.96 Å, it would cause increase in compressive strain on the films as the substrate lattice constants decreases; (STO substrates: $a_c$ ~3.90 Å; LSAT: $a_c$ ~3.88 Å; LAO: $a_c$ ~3.78 Å).

X-ray diffraction (XRD) measurements of the SIRO films on these substrates show the crystalline (00$l$)$_C$ peak without any impurity or additional peaks (Fig. 1a). To further characterize the crystallinity of the films, $\omega$ scans (rocking curve) were performed; in the results (Fig. 1b) around the (002)$_C$ peak of SIRO films, the center positions of full width at half maximum (FWHM) coincided and made to zero value. It is seen from the figure that the FWHM is small (∼ 0.06°) for films on STO and LSAT but becomes broad (∼ 1.65°) for films on LAO. The dense twinning of the LAO substrate is probably responsible for degrading the surface quality of film on LAO substrate; this kind of feature has also been seen in pure $SrRuO_3$ film grown on LAO substrate.[24] To confirm more about sample quality, especially in the SIRO/STO film, (as magnetotransport and magnetization measurements were done in this film based on substrate's popularity; shown later) we used secondary ion mass spectrometry (SIMS) depth profile analysis to check the distribution of the cationic elements within the films. The distributions of Sr, Ir, and O were uniform over the whole film thickness at the resolution level of SIMS (Fig. 1c). Atomic force microscopy (AFM) revealed that all surfaces were flat with a roughness of ∼1-2 nm (inset, Fig. 1c). We have also measured the strain in the SIRO film grown on STO substrate utilizing reciprocal space mapping (RSM); a pathway to extract the inplane and out-of-plane lattice constants as well as the amount of strain by mapping of symmetric and asymmetric Bragg peaks. The RSM result around the inplane $(103)_C$ reflection peak of the SIRO film (Fig. 1d) show that the film and substrate are locked and the strain in the film is determined by the underlying substrate. From the RSM data, the calculated out-of-plane lattice constant of the SIRO film was ∼4.01 Å, which confirms overall inplane compression with out-of-plane lattice expansion of the film.

Measured electrical resistivity of the SIRO films deposited on the substrates under the same deposition conditions were affected by temperature (Fig. 2a). The transport properties of the SIRO films were sensitive to the underlying substrates, as transport in both parent compounds ($SrRuO_3$, $SrIrO_3$) depends on underlying substrates.[7,10] In all three cases, the SIRO displays a metallic characteristic (decrease of resistance with decrease in temperature), with a metal-to-insulator transition (MIT) at a certain temperature $T_{MIT}$ and then insulating characteristics below it (increase of resistance with decrease in temperature), down to 10 K. It is clearly seen that $T_{MIT}$ varied systematically from ~80 K (for STO), ~90 K (for LSAT) and ~100 K (for LAO) depending on the underlying substrate (Fig. 2a). This trend suggests that the strength of disorder in these films changes as the lattice mismatch between film and substrate increases, due to thermodynamic tendency to minimize the strain energy. In parent SrIrO$_3$ film on STO, we also observed a MIT at $T$=50 K and as well as more compressive strained films show fully insulating nature in resistivity related with localization effect.[10,25] To perform quantitative analysis of each resistivity curve and gain further insights, at first in the metallic regime, we attempted to fit the temperature dependence of the resistivity and fitted with typical formula: $\rho = \rho_0 + AT^{3/2}$, (Fig. 2b) where $\rho_0$ is the residual resistivity which signifies scattering due to disorder, and $A$ is the temperature coefficient, which represent the scattering strength. Here to be stressed that this fit was similar to metallic resistivity fitting for parent SrIrO$_3$ on STO substrate, where we also observed a $T^{3/2}$ power law dependence, but was not similar to the $T^2$ (at $T < T_C$) or linear $T$ dependence (at $T > T_C$) of SrRuO$_3$ film.[7,10] Generally, contribution of the incoherent part of electron magnon scattering to resistivity shows $T^{3/2}$ dependent at *low temperature* as found by Mills *et al*.[26] Our resistivity data cannot be interpreted using the above conventional theory because the $T^{3/2}$ power law was observed over a rather high range of ~80 K $\leq T \leq$ 300 K. Self consistent renormalization theory based on *spin fluctuations* also produces a $T^{3/2}$ power law in resistivity.[27] This explanation can also be excluded in our system because it does not show any kind of long-range magnetic ordering (will be shown later). But the resistivity fitting shows $A$ ~4×10$^{-8}$ Ω·cmK$^{-1.5}$, which is close to the value of ~3-5×10$^{-8}$ Ω·cmK$^{-1.5}$, which is often found in other strongly correlated electron systems such as Na$_{0.5}$CoO$_2$ and CaVO$_3$.[28]

As the system goes through the transition to Mott Hubbard regime from the itinerant side or vice versa, the quasi-particle scenario breaks down and non-Fermi liquid resistivity ($\rho \propto T^{3/2}$) was predicted to be associated with scattering between localized electrons by locally cooperative bond-length fluctuations and itinerant electrons. Moreover in the insulating region below $T_{MIT}$ (Fig. 2c), $\rho$ of the SIRO films scaled with the variable range hopping (3D-VRH) model in three dimensions (i.e. $ln\ \sigma \propto 1/T^{1/4}$).[29] In the insulating regime, the Anderson localization effect occurs with the possibility of magnetic clustering effect of Ru and Ir moments. Transport is then only possible by hopping of charge carriers between localized states; this restriction suggests that the increase in $\rho$ at $T < T_{MIT}$ is a result of Anderson localization due to disorder caused by the random distribution of Ir ions on the Ru sites and vice-versa. This increase in disorder strength with increasing compressive strain is quite possible and would increases $T_{MIT}$ in samples grown under large lattice mismatch as the compressive strain suppresses the hopping probability of charge carriers.[10,25] The possibility that the increase in $\rho$ might be due to the magnetic impurity Kondo effect has been excluded because this Kondo effect can only be observable at the *low temperature* limit and our temperature limit is quite high with respect to that limit.[30] Although identifying the exact mechanism for the MIT and systematic variation of $T_{MIT}$ on different substrates requires further investigations on microscopic level, the present resistivity data demonstrate that MIT in SIRO is quite sensitive to external perturbations. Also we may emphasize that the residual resistivity ratio ($RRR \sim 1$), is rather small, so the systems are be regarded as bad metals, as is parent SrIrO$_3$.[10,25]

Now turning into another aspect of how applied external magnetic field ($B$) affects the resistivity (*viz.* magnetoresistance, MR), would be an interesting phenomenon to observe. At low temperature, SrRuO$_3$ shows negative MR. When subjected to a magnetic field, Ru moments become well-ordered along its easy axis and thus reduce the scattering, so MR becomes negative; this is a well-established physical effect in ferromagnetic metals. In contrast, SrIrO$_3$ shows positive MR when subjected to a magnetic field and this response can be explained to result from the Lorentz contribution for fully metallic films.[10] The three dimensional SIrO$_3$ films with increase in resistivity also show positive MR, to which both correlation and disorder make important contributions.[10] Despite the origin, a three dimensional SrRuO$_3$ shows *negative* MR whereas a three dimensional SrIrO$_3$ film shows *positive* MR. SrRuO$_3$ also has much larger MR values than SrIrO$_3$; this difference implies that the magnetic field has much stronger effect on Ru spins due to its ferromagnetic nature than its counterpart of possible paramagnetic moments of Ir spins.

Therefore, we measured magnetoresistance with the applied magnetic field oriented perpendicular to the film surface. MR of SIRO at $T=10$ K was negative for samples grown on all three substrates (Fig. 3). At a glance it can be said that at low temperature, the applied magnetic field mainly aligns the Ru spins (either ferromagnetic-like or some kind of local ordering; glassy like) and thereby leading to a decrease of electron-spin scattering and to a negative MR. As said earlier, 4$d$ itinerant ferromagnetic Ru moments have much stronger effect than 5$d$ Ir moments on applying magnetic field, so MR remains negative, although its magnitude decreases due to competitive randomness of spin interactions.

To check how this competition is affected by temperature, we took one sample (film on STO because it is the most popular substrate used for thin film growth) and measured the temperature dependence of MR (Fig. 4a). As temperature increased, MR changed sign and became positive. While increasing the temperature further, MR diminishes as seen in both parent materials also. Clearly there is a sigh change in MR above ∼25 K (Fig. 4b). Plots of $\rho$ vs. $T$ at magnetic strengths of 0 T and 9 T crossed at $T \geq 25$ K (Fig. 4b, inset). Moreover negative MR was found to be $\propto B^{1/2}$ (Fig. 4c); consistent with localization picture. [31] In contrast, positive MR follows $B^2$ power law dependence (Fig. 4d); consistent with normal transport behavior. This observation suggests that as $T$ increases, the applied magnetic field mainly localizes an increasing number of 5$d$ moments by enhancing the exchange interaction between localized moments and itinerant moments, thereby resulting in formation of magnetic polarons, and consequent appearance of positive MR. Switching of MR observed in other materials has also often been related to magnetic scattering.[32-34]

To get an intuition about magnetic ordering in SIRO/STO, magnetization measurement of SIRO/STO film was performed; this film did not show any long-range magnetic ordering in the measured temperature range; 10 K ≤ $T$ ≤ 300 K. This result is consistent with the lack of any twisting in the resistivity curve (Fig. 2a); this lack of twisting indicates the absence of long-range magnetic order. In the localization picture it's quite natural that the randomness of Ir paramagnetic moments due to its extended 5$d$ orbitals affects the local environment of Ru atoms and leads to suppression of its long-range ferromagnetic state. But even in this scenario, local ordering can occur at low $T$ and be attributed to local magnetic ordering (spin glass type; $T_{SG}$) as seen at $T \leq 25$ K (Fig. 5a). Although the moment of the film was very low (as expected for a system containing a 5$d$ element that has weak correlation) even at low temperature (with respect to parent SrRuO$_3$) and close to the noise level, but the clearly observable cusp at $T \sim 25$ K establishes the appearance of glassy like behavior. Similar features of glassy ordering were also been reported in other systems. [35-37] Possible origin of glassy magnetic ordering due to substrate has been excluded, because magnetization measurements on STO substrate show purely diamagnetic behavior. Therefore the coexistent tendency of localization-induced insulating nature in transport (Fig. 2c) and glassy ordering (Fig. 5a) validates with the hypothesis that Anderson localization effect strongly stabilizes spin glassy behavior while Mott localization tends to suppress it.[38] Surprisingly, we did not observe any glassy features in SIRO films grown on LSAT or LAO substrates (Fig. 5b and 5c). Any long-range magnetic ordering or low-temperature spin-glass behavior might have disappeared with much small percentage of Ir doping on SrRuO$_3$. This explanation is plausible because the evolution of the magnetic phase diagram is different for films grown on different substrates, because they have different amounts of strain, and increasing disorder (which increases with the lattice mismatch) affects the magnetization. On this basis, the origin of negative MR at low $T$ for higher strained films could be due to localization effect associated with the increased random distribution of Ir ions on the Ru sites and vice-versa due to increased strain. Assuming the above scenario stands the test of time, to shed the light more, perhaps the most important step would be to obtain detailed information on microscopic level. Currently, X-ray magnetic circular dichroism (XMCD) measurements are ongoing to investigate the exact nature of magnetic ordering and testify the role of each transition metal element.

## IV. CONCLUSIONS

We grew epitaxial SrIr$_{0.5}$Ru$_{0.5}$O$_3$ thin films on various lattice-mismatched substrates (e.g. SrTiO$_3$ (001), (LaAlO$_3$)$_{0.3}$(Sr$_2$TaAlO$_6$)$_{0.7}$ (001) and LaAlO$_3$ (001)). The films showed systematic variation in metal-to-insulator transition temperature $T_{MIT}$ in resistivity $\rho$ ($T_{MIT}\sim$ 80 K on SrTiO$_3$, $\sim$90 K on (LaAlO$_3$)$_{0.3}$(Sr$_2$TaAlO$_6$)$_{0.7}$ and $\sim$100 K on LaAlO$_3$); this variation is associated with disorder caused by large lattice mismatch. The entire metallic regimes of $\rho$ followed a $T^{3/2}$ power laws on all three films, whereas insulating regimes were consistent with a three dimensional variable range hopping model. Furthermore, all the films show negative magnetoresistance (MR) at $T$=10 K. MR switched from negative to positive as temperature increased above $\sim$25 K for SrIr$_{0.5}$Ru$_{0.5}$O$_3$ film on SrTiO$_3$ substrate. This temperature was associated with the signature of spin glassy ordering ($T_{SG}\sim$25 K), consistent with localization theory. Overall, switching of MR may provide new physical significances which can be used for novel technological applications in which coupling of two different ground states persist. In general, disorder related (quantum) phenomena have fundamental rich physical aspects [31] and above results are elucidated following the basic principle from disordered physics perspectives. However, above explanations may not be conclusive enough and requires detailed microscopic analysis, experimentally as well as theoretically for deeper understandings, considering the role of each and every element. So, in future it is highly desirable to grow fully atomically ordered layer-by-layer thin films, especially SrIrO$_3$/SrRuO$_3$ superlattices, to further unveiling of new emerging phenomena.

## ACKNOWLEDGMENTS

We would like to thank Prof. B. I. Min for informative discussions. Authors would like to thank NINT at POSTECH for SIMS measurements and KBSI Daegu for RSM measurements.

This work was supported by the National Research Foundation via SRC at POSTECH (2011-0030786).

**REFERENCES**

1. J. B. Goodenough, *Localized to Itinerant Electronic Transition in Perovskite Oxides* (Springer Series in Structure and Bonding, Vol. 98, Berlin: Springer, 2001)

2. J. B. Goodenough and J.-S. Zhou, Chem. Mater. **10**, 2980 (1998). (10.1021/cm980276u)

3. M. Imada, A. Fujimori, and Y. Tokura, Rev. Mod. Phys. **70**, 1039 (1998). (http://dx.doi.org/10.1103/RevModPhys.70.1039)

4. S. J. Moon, H. Jin, K. W. Kim, W. S. Choi, Y. S. Lee, J. Yu, G. Cao, A. Sumi, H. Funakubo, C. Bernhard, and T. W. Noh, Phys. Rev. Lett. **101**, 226402 (2008). (http://dx.doi.org/10.1103/PhysRevLett.101.226402)

5. B. J. Kim, H. Ohsumi, T. Komesu, S. Sakai, T. Morita, H. Takagi, and T. Arima, Science **323**, 1329 (2009). (10.1126/science.1167106)

6. W. W. -Krempa, G. Chen, Y.-B. Kim, and L. Balents, Condens. Matter Phys. **5**, 2014 (2013). (10.1146/annurev-conmatphys-020911-125138)

7. G. Koster, L. Klein, W. Siemons, G. Rijnders, J. S. Dodge, C.-B. Eom, D. H. A. Blank, and M. R. Beasley, Rev. Mod. Phys. **84**, 253 (2012) (http://dx.doi.org/10.1103/RevModPhys.84.253) and references therein.

8. J. G. Zhao, L. X. Yang, Y. Yu, F. Y. Li, R. C. Yu, Z. Fang, L. C. Chen, and C. Q. Jin, J. Appl. Phys. **103**, 103706 (2008).(http://dx.doi.org/10.1063/1.2908879)

9. F. –X. Wu, J. Zhou, L. Y. Zhang, Y. B. Chen, S.-T. Zhang, Z. –B. Gu, S. –H. Yao, and Y. -F. Chen, J. Phys.: Condens. Matter **25**, 125604 (2013).(doi:10.1088/0953-8984/25/12/125604)

10. A. Biswas, K. -S. Kim, and Y. H. Jeong, J. Appl. Phys. **116**, 213704 (2014). (http://dx.doi.org/10.1063/1.4903314)

11. K. Yamaura, D. P. Young, and E. T.-Muromachi, Phys. Rev. B **69**, 024410 (2004). (http://dx.doi.org/10.1103/PhysRevB.69.024410)

12. H. Nakatsugawa, E. Iguchi, and Y. Oohara, J. Phys.: Condens. Matter **14**, 415 (2002). (doi:10.1088/0953-8984/14/3/311)

13. T. He, Q. Huang, and R. J. Cava, Phys. Rev. B **63**, 024402 (2000). (http://dx.doi.org/10.1103/PhysRevB.63.024402)

14. G. Cao, S. McCall, M. Shepard, J.E. Crow, and R.P. Guertin, Phys. Rev. B **56**, 321 (1997). (http://dx.doi.org/10.1103/PhysRevB.56.321)

15. W. Tong, F. –Q. Huang, and I. -W. Chen, J. Phys.: Condens. Matter **23**, 086005 (2011). (doi:10.1088/0953-8984/23/8/086005)

16. M. Bremholm, C. K. Yim, D. Hirai, E. Climent-Pascual, Q. Xu, H. W. Zandbergen, M. N. Ali and R. J. Cava, J. Mater. Chem., **22**, 16431 (2012). (10.1039/C2JM32558F)

17. I. Qasim, B. J. Kennedy and M. Avdeev, J. Mater. Chem. A, **1**, 3127 (2013). (10.1039/C3TA00540B)

18. S. Y. Jang, H. Kim, S. J. Moon, W. S. Choi, B. C. Jeon, J. Yu and T. W. Noh, J. Phys.: Condens. Matter **22**, 485602 (2010). (doi:10.1088/0953-8984/22/48/485602)

19. J. Nichols, J. Terzic, E. G. Bittle, O. B. Korneta, L. E. De Long, J. W. Brill, G. Cao, and S. S. A. Seo, Appl. Phys. Lett. **102**, 141908 (2013). (http://dx.doi.org/10.1063/1.4801877)

20. A. Biswas, P. B. Rossen, C.-H. Yang, W. Siemons, M.-H. Jung, I. K. Yang, R. Ramesh, and Y. H. Jeong, Appl. Phys. Lett. **98**, 051904 (2011). (http://dx.doi.org/10.1063/1.3549860)

21. F. Sanchez, C. Ocal, and J. Fontcuberta, Chem. Soc. Rev. **43**, 2272 (2014). (10.1039/C3CS60434A)

22. J. M. Longo, J. A. Kafalas, and R. J. Arnott, J. Solid State Chem. **3**, 174 (1971). (doi:10.1016/0022-4596(71)90022-3)

23. C. B. Eom, R. J. Cava, R. M. Fleming, Julia M. Phillips, R. B. vanDover, J. H. Marshall, J. W. P. Hsu, J. J. Krajewski, W. F. Peck Jr., Science, **258**, 1766 (1992). (10.1126/science.258.5089.1766)

24. J. C. Jiang and X. Q. Pan, J. Appl. Phys. **89**, 6365 (2001). (http://dx.doi.org/10.1063/1.1368160)

25. J. H. Gruenewald, J. Nichols, J. Terzic, G. Cao, J. W. Brill and S. S.A. Seo, J. Mater. Res. **29**, 2491 (2014). (http://dx.doi.org/10.1557/jmr.2014.288)

26. D. L. Mills, A. Fert, and I. A. Campbell, Phys. Rev. B **4**, 196 (1971). (http://dx.doi.org/10.1103/PhysRevB.4.196)

27. T. Moriya, *Spin Fluctuations in Itinerant Electron Magnetism* (Springer Series in Solid State Sciences 56, Berlin: Springer, 1985).

28. F. Rivadulla, J. S. Zhou, and J. B. Goodenough, Phys. Rev. B **67**, 165110 (2003). (http://dx.doi.org/10.1103/PhysRevB.67.165110)

29. N. F. Mott, *Metal-Insulator Transitions* (Taylor & Francis Ltd., London, 1974).

30. L. Kouwenhoven and L. Glazman, Phys. World **14**, No. 1, 33–38 (2001).

31. P. A. Lee and T. V. Ramakrishnan, Rev. Mod. Phys. **57**, 287 (1985). (http://dx.doi.org/10.1103/RevModPhys.57.287)

32. X. –H. Li, Y. –H. Huang, Z. –M. Wang, and C. –H. Yan, Appl. Phys. Lett. **81**, 307 (2002). (http://dx.doi.org/10.1063/1.1491281)

33. R. Mahendiran and A. K. Raychaudhuri, Phys. Rev. B 5**4**, 16044 (1996). (http://dx.doi.org/10.1103/PhysRevB.54.16044)

34. H. Li, Y. Xiao, B. Schmitz, J. Persson, W. Schmidt, P. Meuffels, G. Roth, and T. Bruckel, Sci. Rep. **2**, 750 (2012). (10.1038/srep00750)

35. C.-J. Cheng, C. Lu, Z. Chen, L. You, L. Chen, J. Wang and T. Wu, Appl. Phys. Lett. **98**, 242502 (2011). (http://dx.doi.org/10.1063/1.3600064)

36. C. R. Wiebe, J. E. Greedan, G. M. Luke, and J.S. Gardner, Phys. Rev. B **65**, 144413 (2002). (http://dx.doi.org/10.1103/PhysRevB.65.144413)

37. R. Palai, H. Huhtinen, J. F. Scott, and R. S. Katiyar, Phys. Rev. B **79**, 104413 (2009). (http://dx.doi.org/10.1103/PhysRevB.79.104413)

38. V. Dobrosavljević, D. Tanaskovic, and A. A. Pastor, Phys. Rev. Lett. **90**, 016402 (2003). (http://dx.doi.org/10.1103/PhysRevLett.90.016402)

**FIGURE CAPTIONS**

**FIG. 1.** (Color Online) (a) X-ray $\theta$–$2\theta$ scan results of SrIr$_{0.5}$Ru$_{0.5}$O$_3$ (SIRO) films of thickness ∼40 nm grown on SrTiO$_3$ (001), LSAT (001) and LaAlO$_3$ (001) substrates (b) Rocking curves ($\omega$ scans) around the (002)$_C$ peak of SrIrO$_3$ films on three substrates are shown. Zero value of $\Delta\omega$ corresponds to the maximum peak position. (c) Secondary ion mass spectrometry (SIMS) reveals homogeneity of the SIRO/STO films (inset; AFM image). (d) Reciprocal space mapping of the (103)$_C$ peak of SIRO/STO reveals that film is fully strained.

**FIG. 2.** (Color Online) (a) Temperature dependence of the resistivity of SrIr$_{0.5}$Ru$_{0.5}$O$_3$ films grown on three different substrates. Metal-to-insulator transitions were observed in all films with $T_{\text{MIT}}$ ∼ 80 K (on STO), ∼90 K (on LSAT), and ∼100 K (on LAO). (b) Resistivity in the metallic regime was fitted with a $T^{3/2}$ power law (300 K > $T$ > $T_{MIT}$). (c) Three-dimensional variable range hopping (3D-VRH) model ($ln\ \sigma \propto 1/T^{1/4}$) was scaled in the insulating regime ($T < T_{MIT}$). (Brown line corresponds to linear fit guided to the eye).

**FIG. 3.** (Color Online) (a) Magnetoresistance (MR) of $SrIr_{0.5}Ru_{0.5}O_3$ films on three substrates. The transverse MR is negative at $T=10$ K for all three films.

**FIG. 4.** (Color Online) (a)-(b) Temperature dependent transverse magnetoresistance (MR) of $SrIr_{0.5}Ru_{0.5}O_3$ film grown on STO substrate is shown. Shaded region shows positive side of MR. As temperature increases, sign of MR switches form negative to positive. Inset shows crossover in resistivity at $T \geq 25$ K with applied magnetic field ($B$) of 9 T. (c)-(d) Negative MR $\propto B^{1/2}$ at $T=10$ K, whereas positive MR $\propto B^2$ at $T=30$ K are shown.

**FIG. 5.** (Color Online) (a)-(c) Zero field cooled (ZFC) and field cooled (FC) magnetization measurements of $SrIr_{0.5}Ru_{0.5}O_3$ films grown on STO, LSAT and LAO substrates with the applied magnetic field of 1000 Oe. Due to the very low overall moment, even noise signal catches in the magnetization curve. Visible cusp at $T \sim 25$ K from ZFC measurement for film grown on STO shows the signature of spin glassy ($T_{SG}$) behavior.

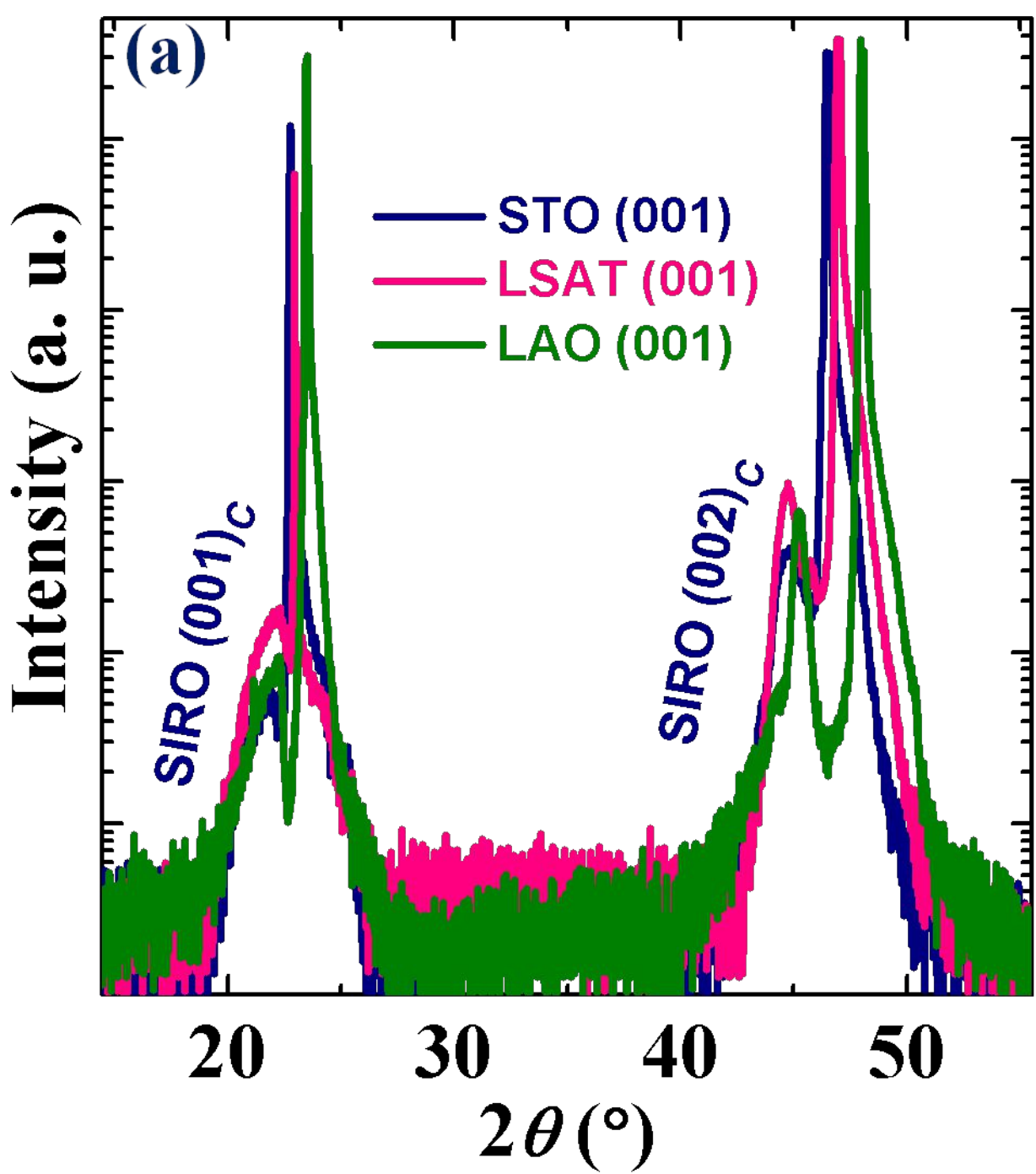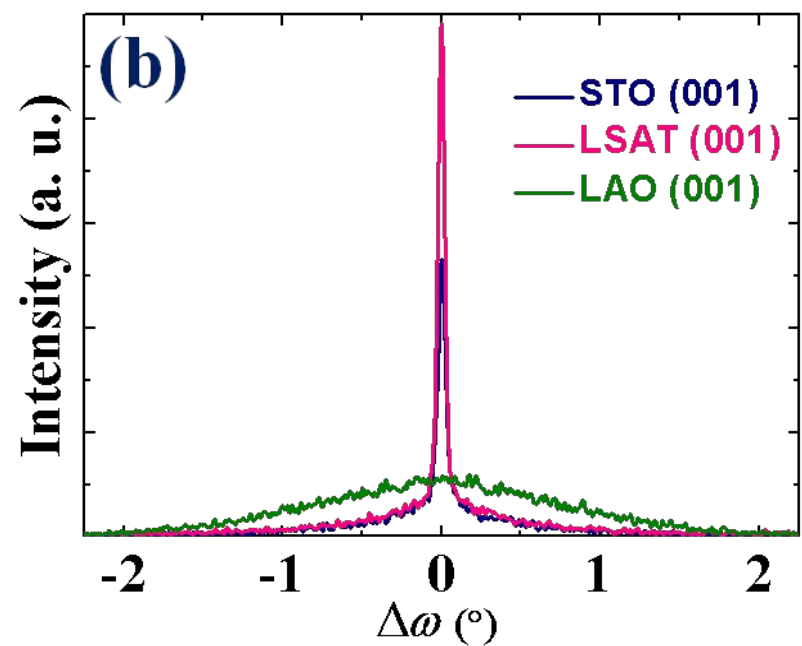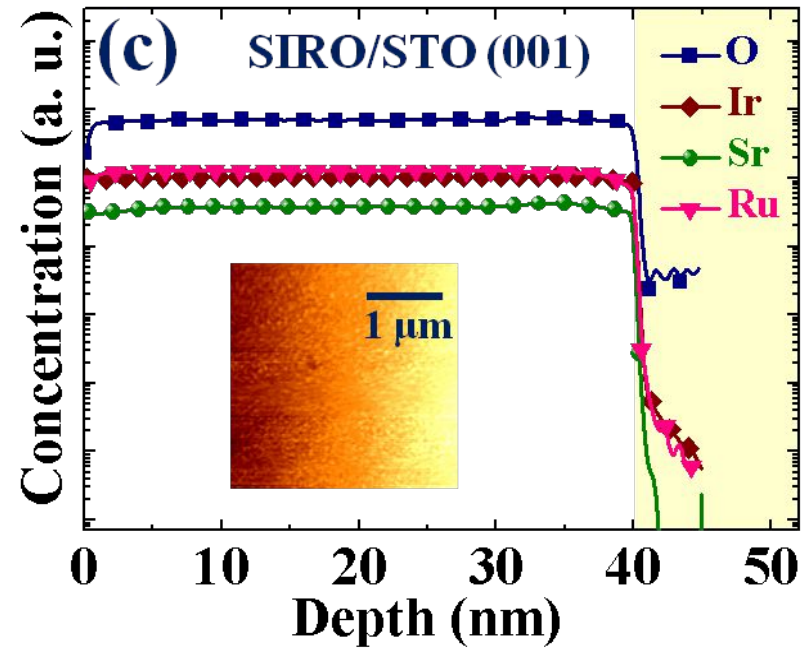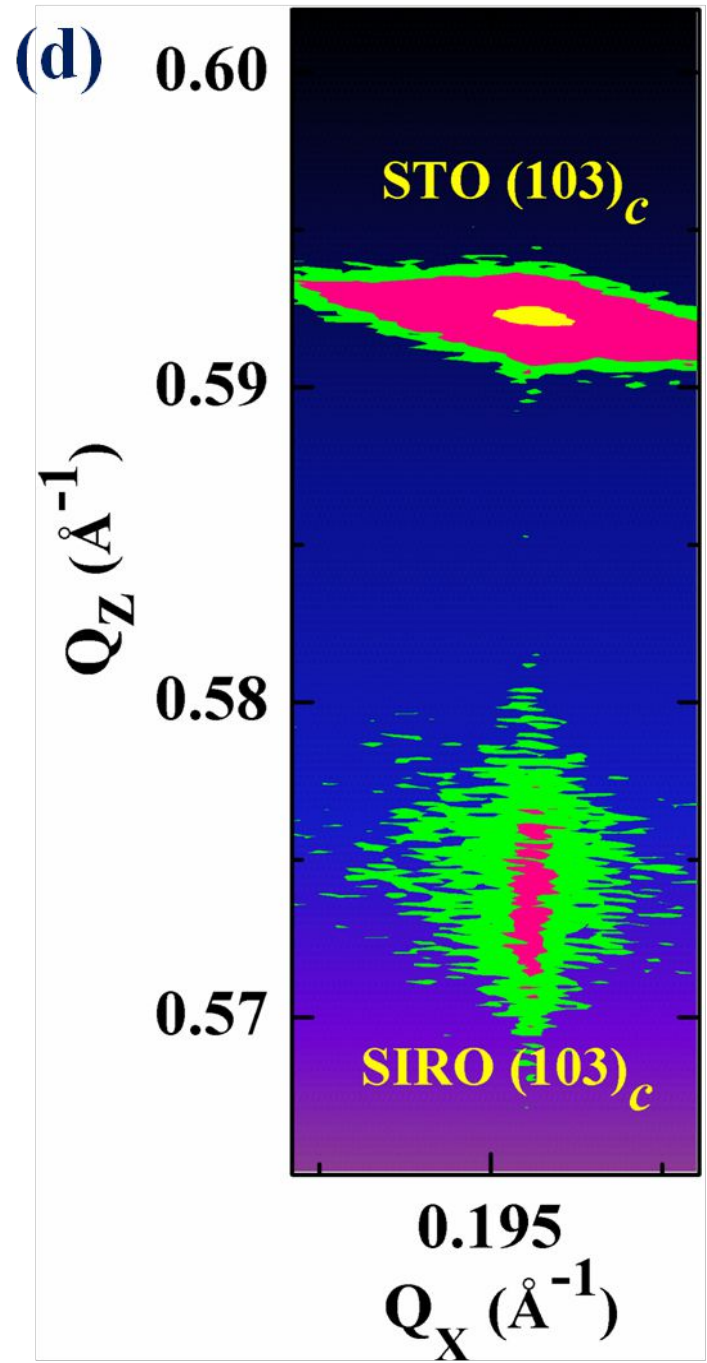

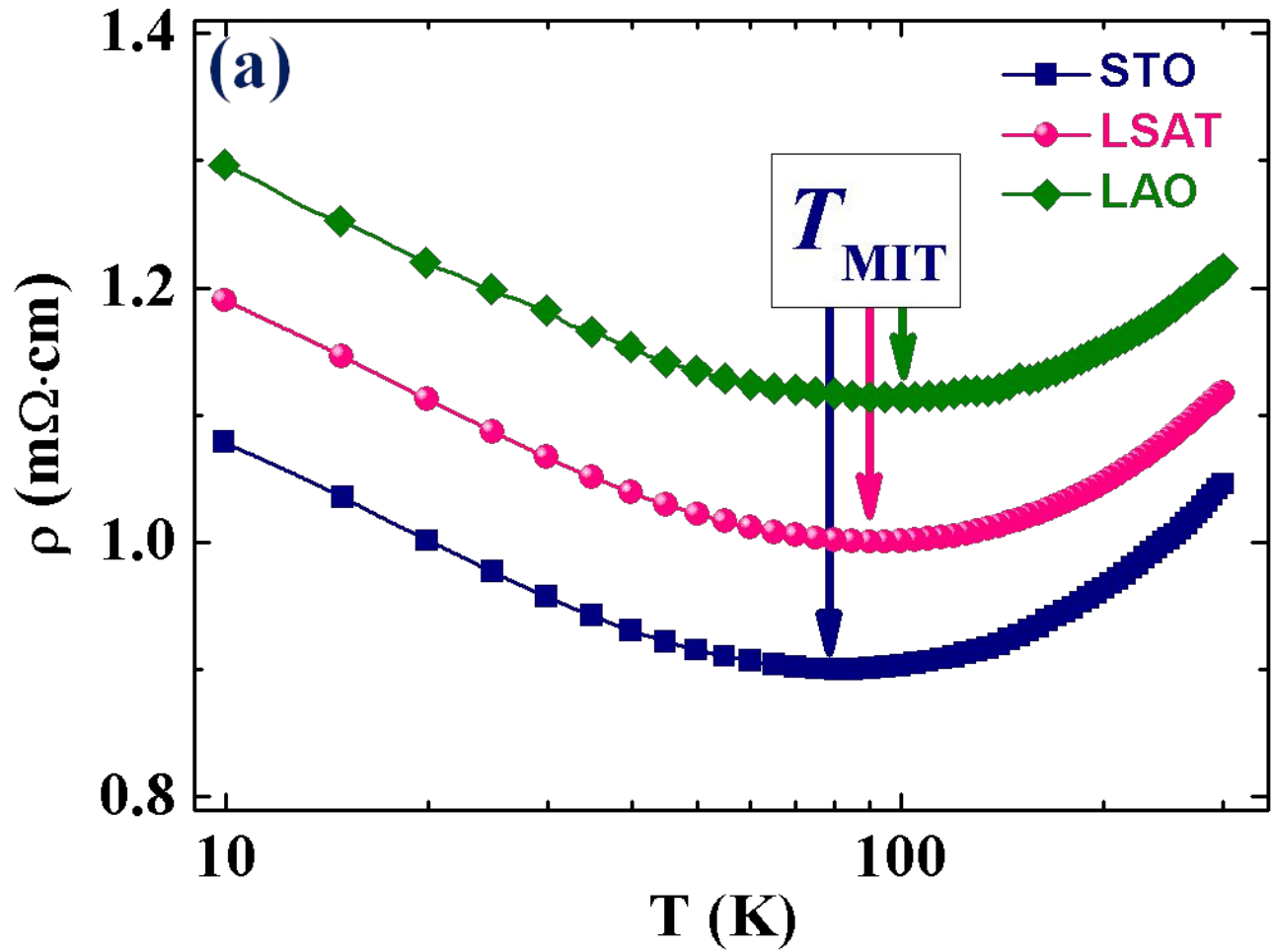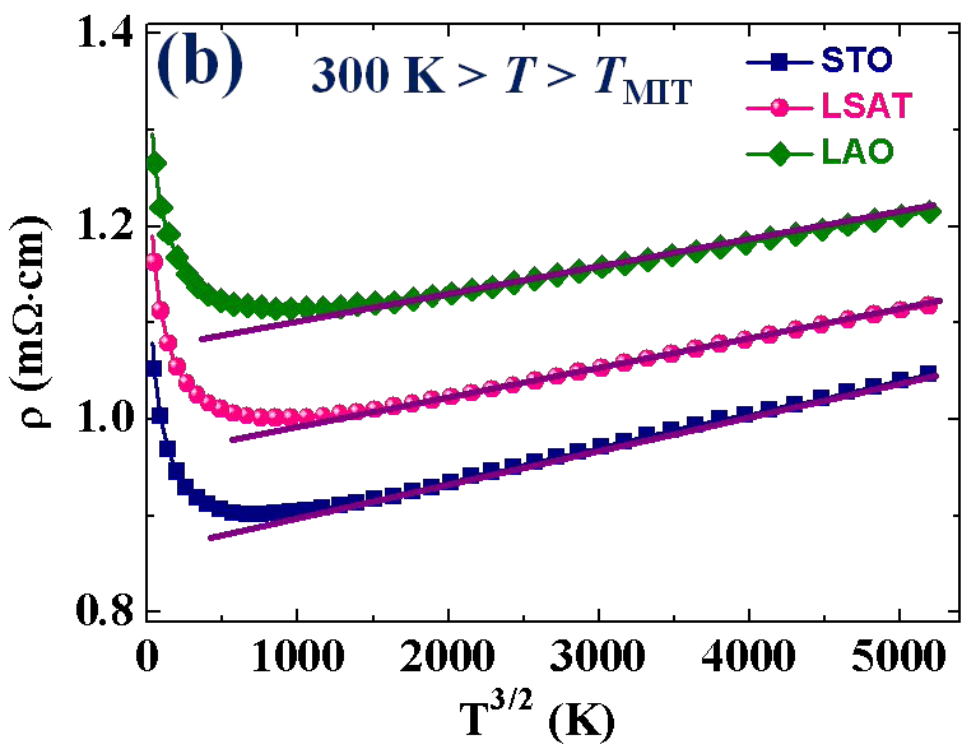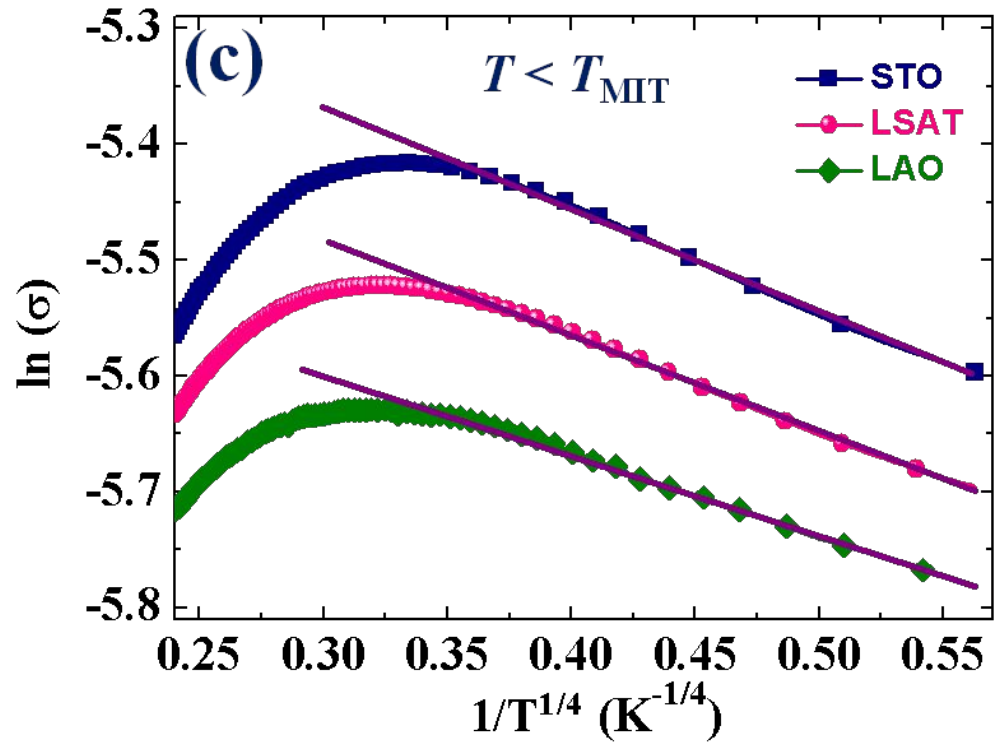

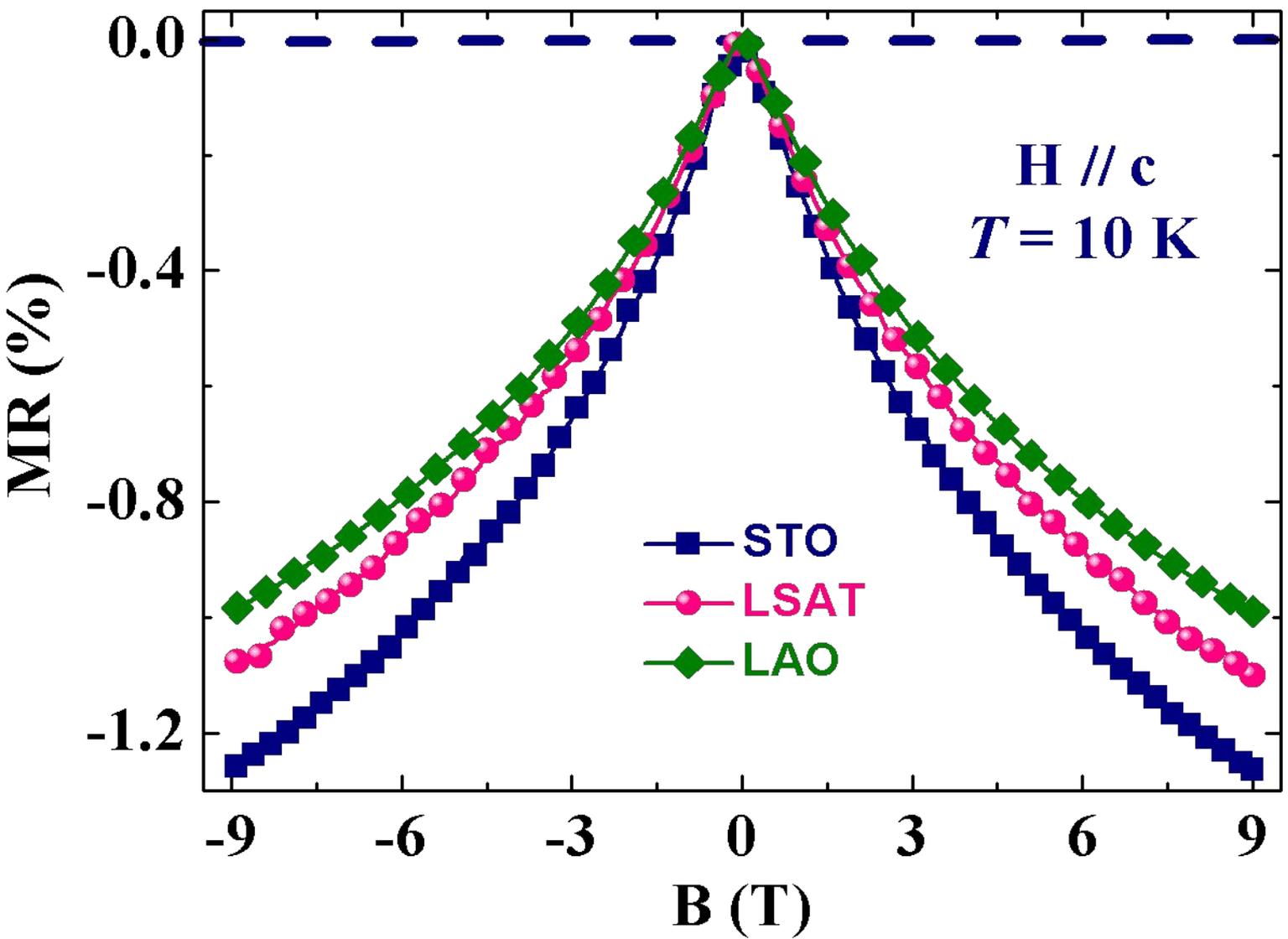

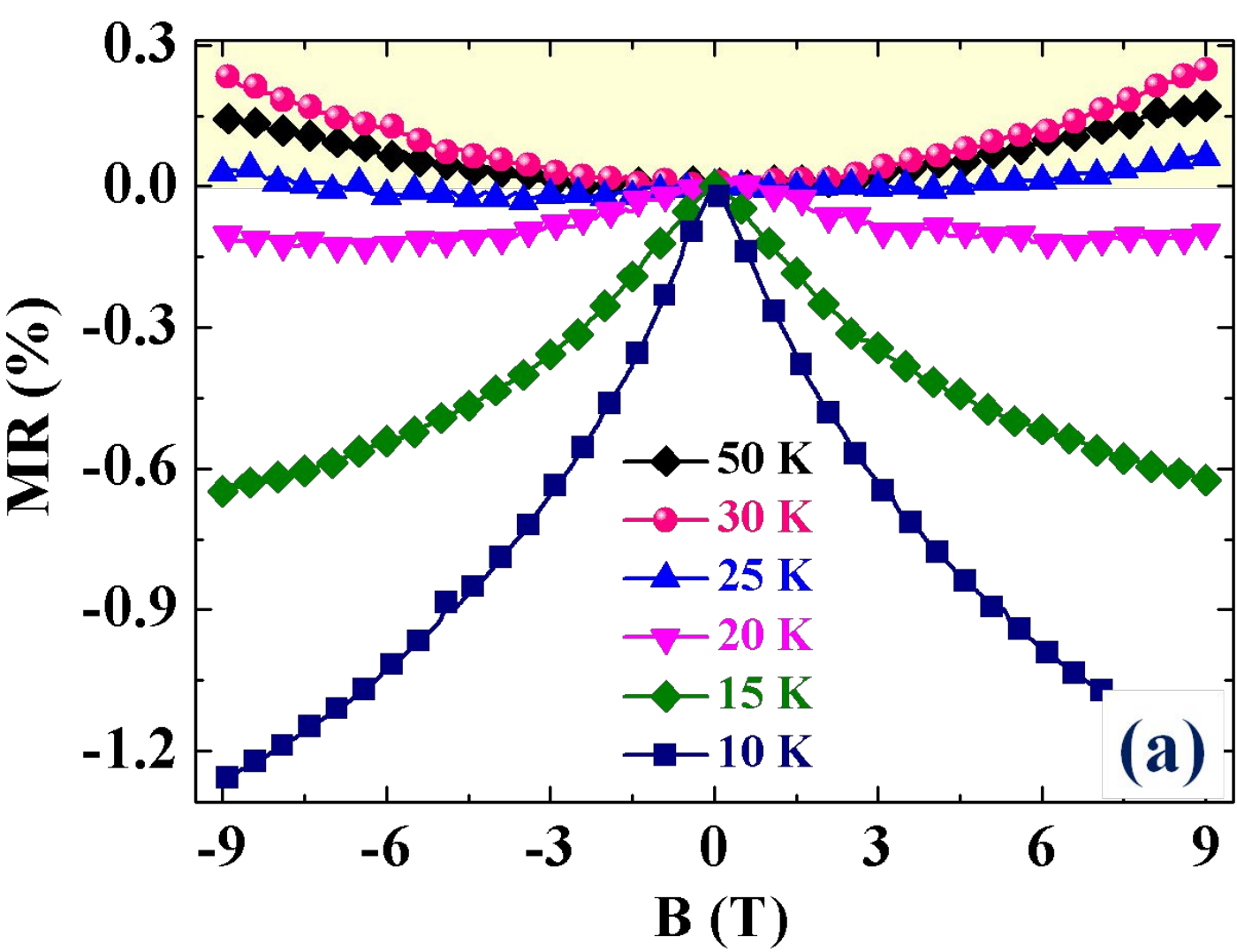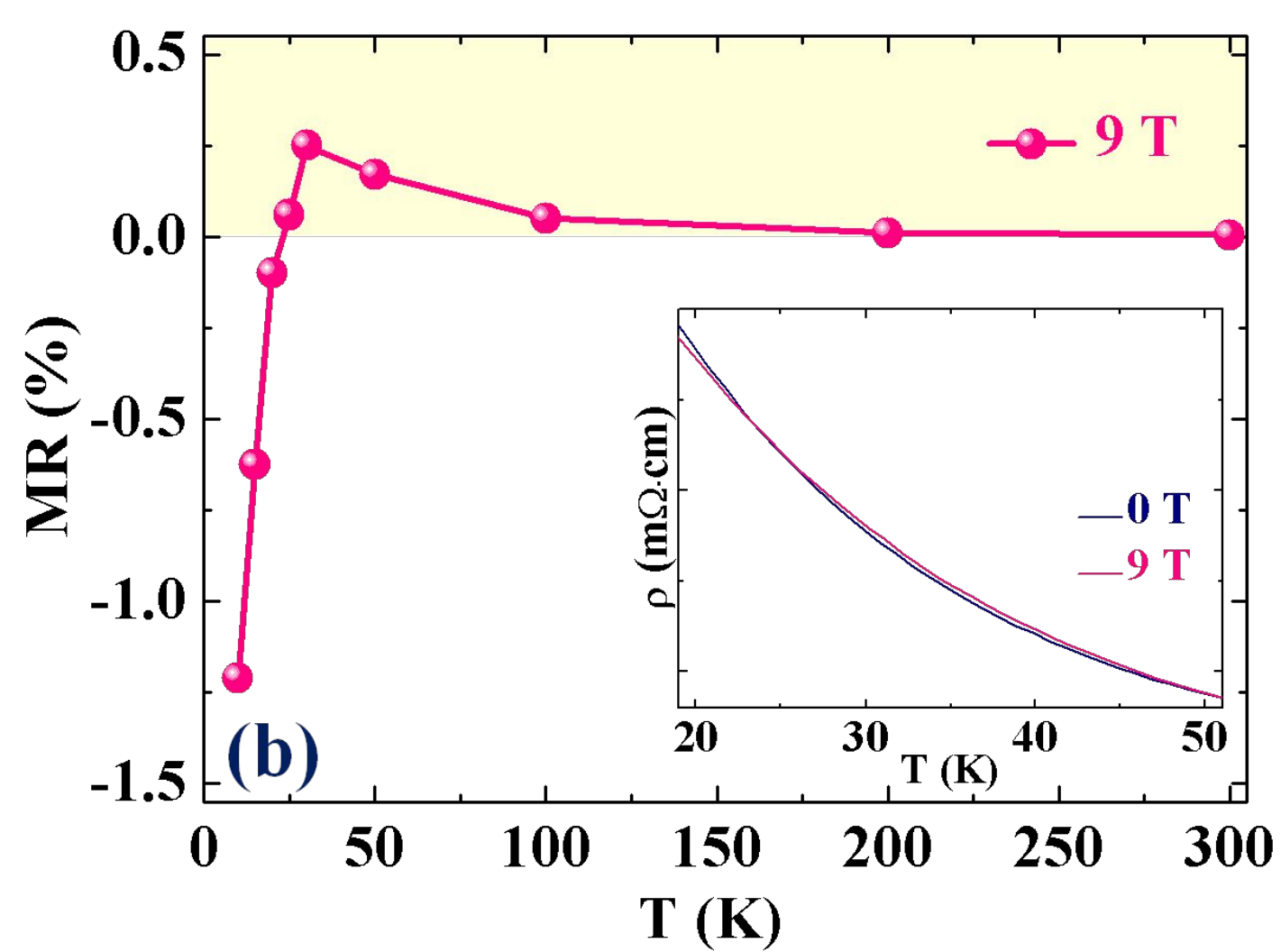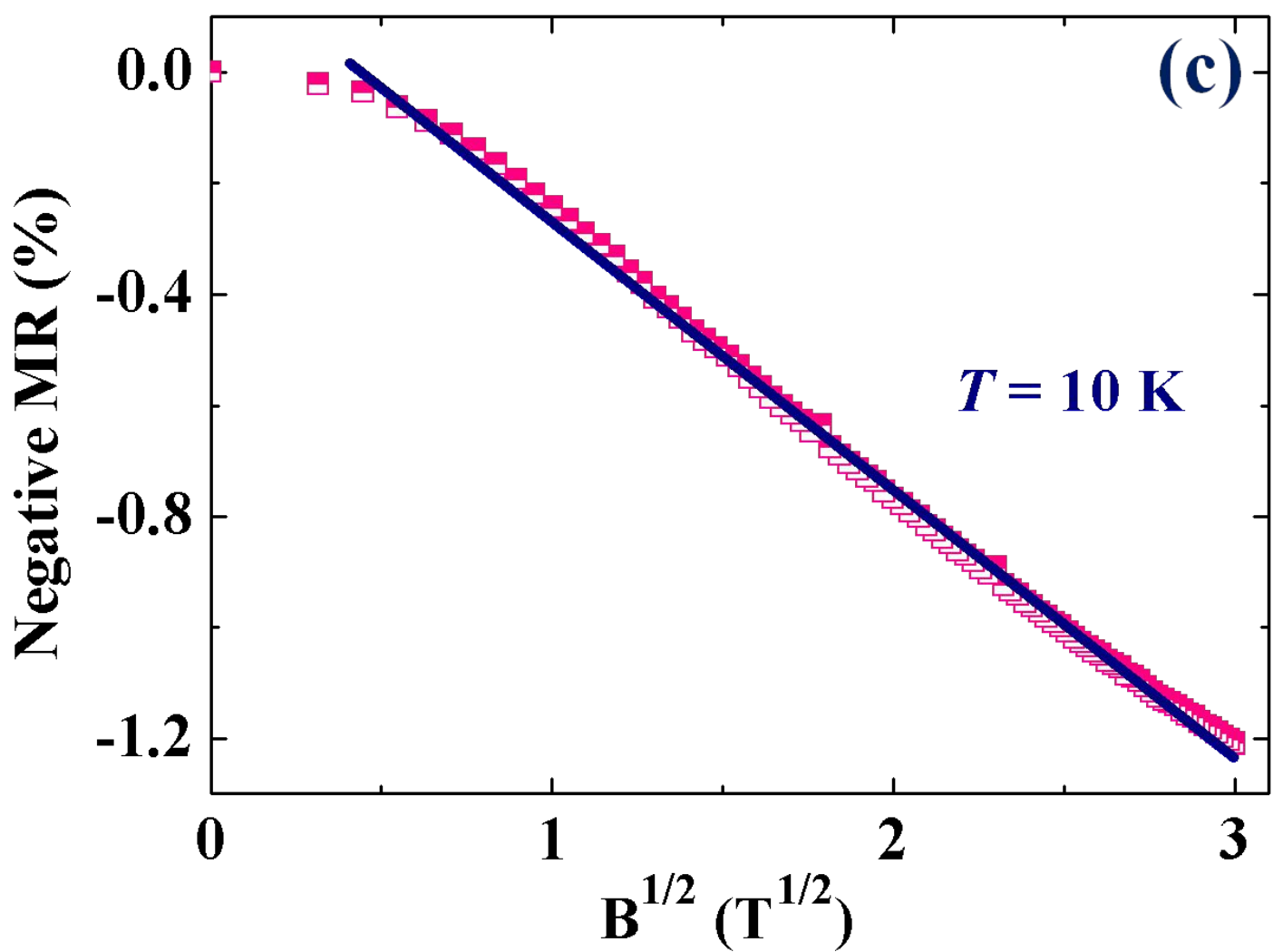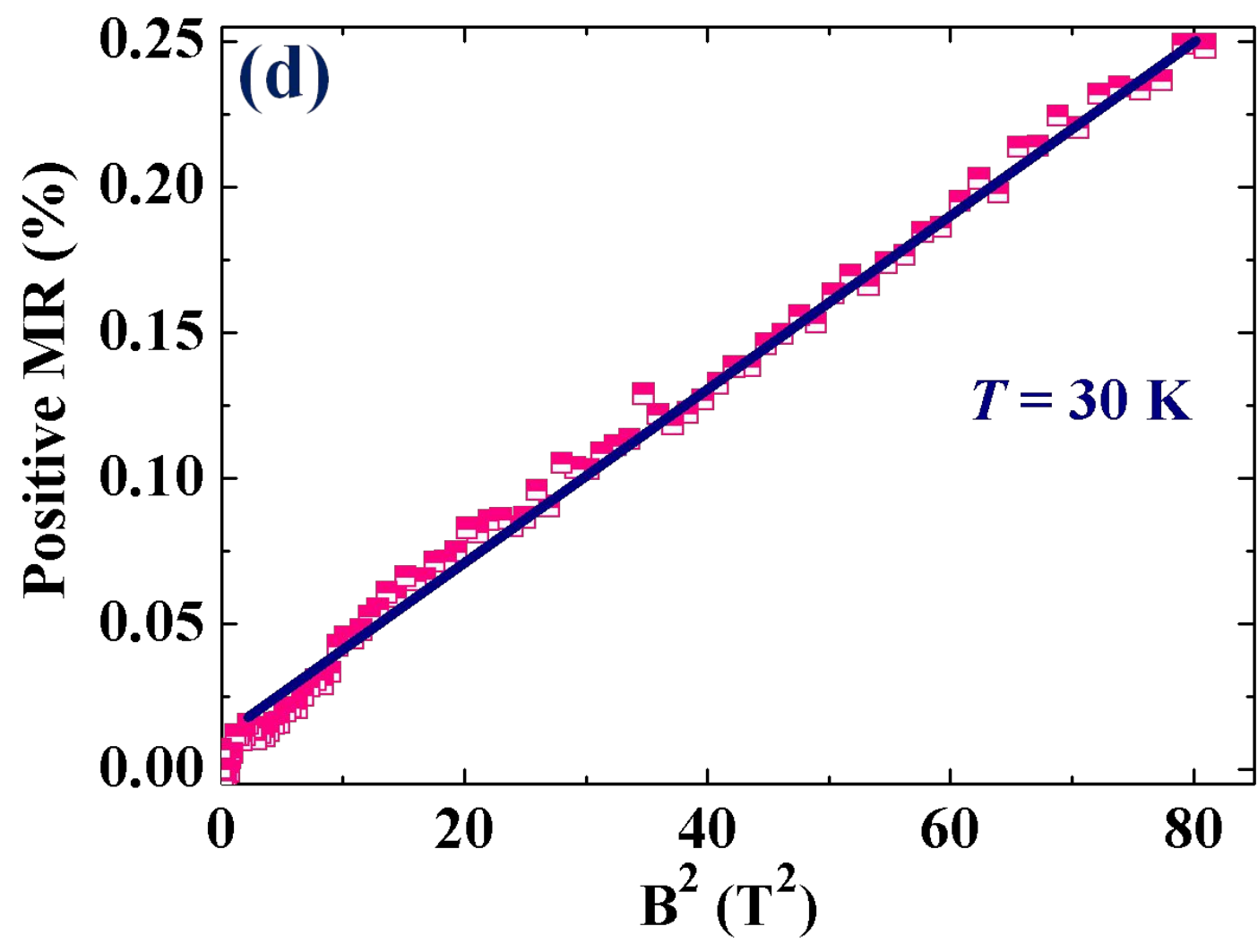

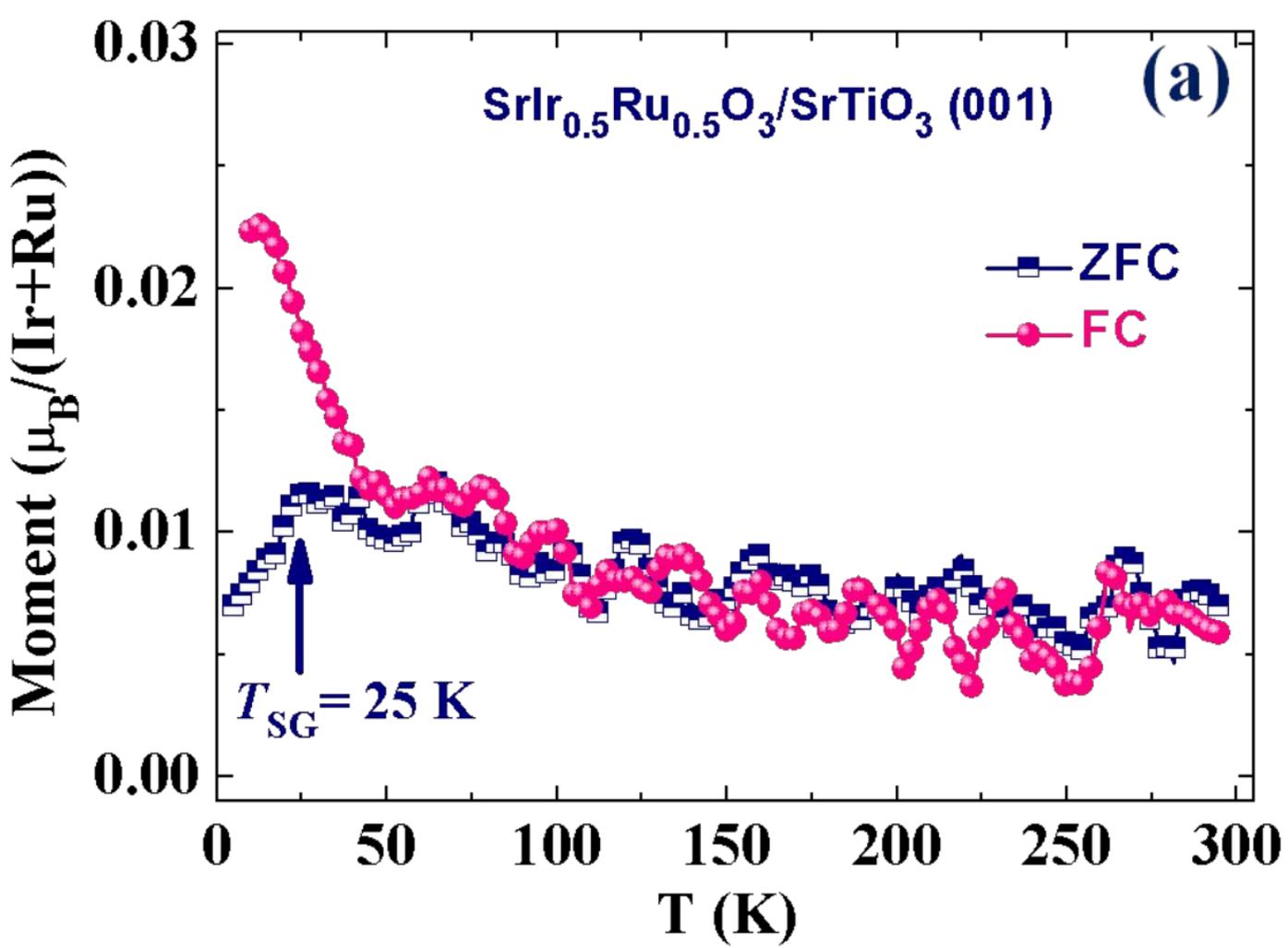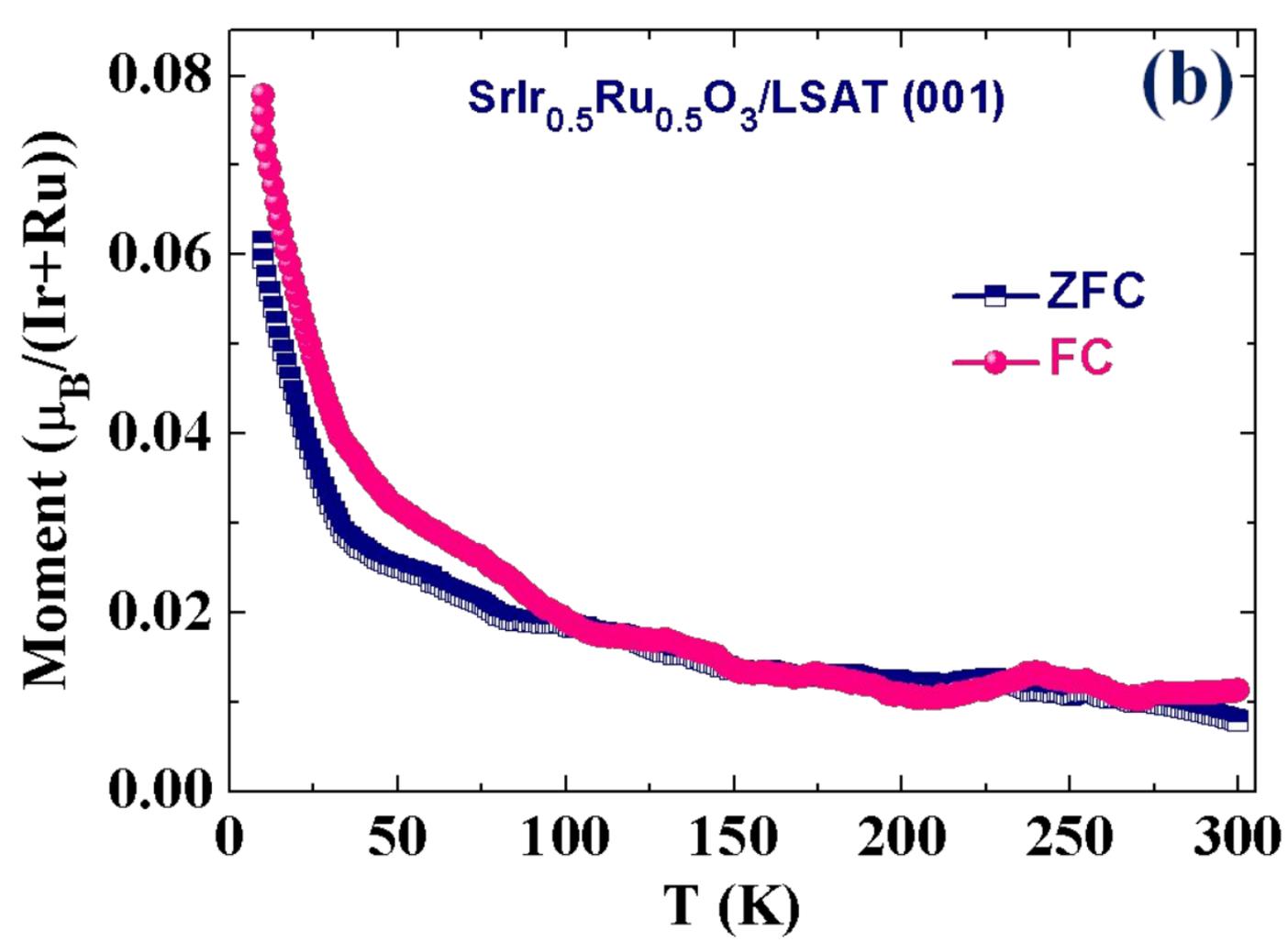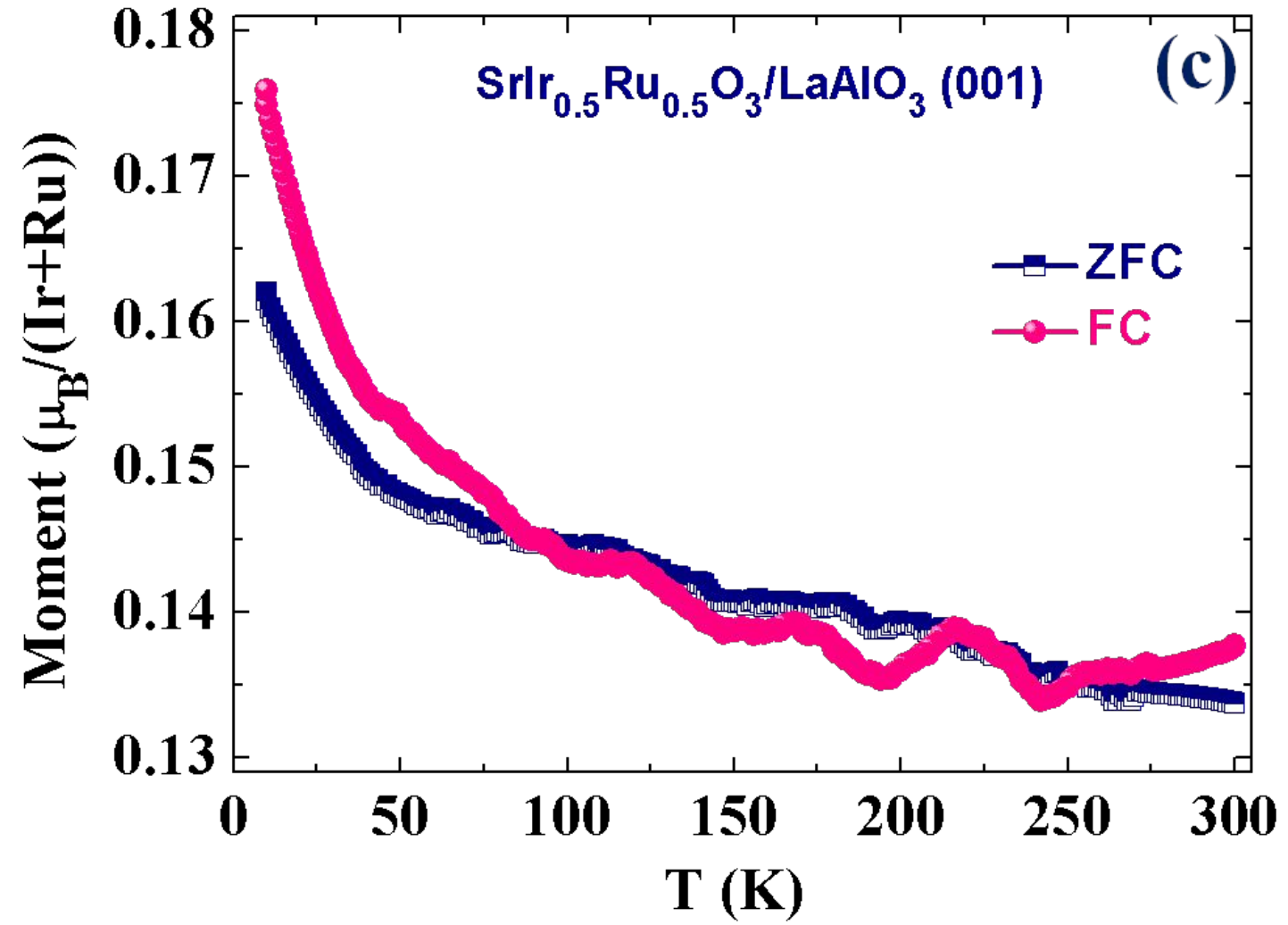